\documentclass{ar-1col-S2O}

\usepackage[numbers]{natbib}
\usepackage{url}
\usepackage{graphicx}
\usepackage{enumitem}
\usepackage{mathtools}
\usepackage{graphicx}   
\usepackage{color}
\usepackage[normalem]{ulem}
\usepackage{caption}
\usepackage{subcaption}
\usepackage{braket}
\usepackage{comment}
\usepackage{bm}
\usepackage{array}
\usepackage{booktabs}
\usepackage{multirow}

\usepackage{amsmath}

\usepackage{booktabs}

\usepackage[breaklinks=true]{hyperref}
\usepackage{breakcites}
\usepackage{algorithm}
\usepackage{algpseudocode}
\usepackage{xcolor}
\usepackage{bm}

\newcommand{\vc}[1]{\mathbf{#1}}
\newcommand{\cl}{\mathrm{ClNO}_2}
\newcommand{\n}{\mathrm{N}_2\mathrm{O}_5}

\newcommand{\nit}{\mathrm{HNO}_3}

\begin{document}
\title{Molecular Insights into Chemical Reactions at Aqueous Aerosol Interfaces}

\author{David T. Limmer,$^{1,2,3,4}$ Andreas W. G{\"o}tz,$^{5}$ Timothy H. Bertram,$^6$ and Gilbert M. Nathanson$^{6,*}$
\affil{$^1$Department of Chemistry, University of California, Berkeley, CA, USA}
\affil{$^2$Kavli Energy NanoScience Institute, Berkeley, CA, USA.}
\affil{$^3$MSD, Lawrence Berkeley National Laboratory, Berkeley, CA, USA}
\affil{$^4$CSD, Lawrence Berkeley, National Laboratory, Berkeley, CA, USA}
\affil{$^5$SDSC, University of California San Diego, La Jolla, CA, USA}
\affil{$^6$Department of Chemistry, University of Wisconsin-Madison, Madison, WI, USA}
\affil{$^*$gmnathan@wisc.edu}}


\date{\today}
\begin{abstract}
Atmospheric aerosols facilitate reactions between ambient gases and dissolved species. Here, we review our efforts to interrogate the uptake of these gases and the mechanisms of their reactions both theoretically and experimentally. We highlight the fascinating behavior of $\n$ in solutions ranging from pure water to complex mixtures, chosen because its aerosol-mediated reactions significantly impact global ozone, hydroxyl, and methane concentrations. As a hydrophobic, weakly soluble, and highly reactive species, $\n$ is a sensitive probe of the chemical and physical properties of aerosol interfaces. We employ contemporary theory to disentangle the fate of $\n$ as it approaches pure and salty water, starting with adsorption and ending with hydrolysis to $\nit$, chlorination to $\cl$, or evaporation. Flow reactor and gas-liquid scattering experiments probe even greater complexity as added ions, organic molecules, and surfactants alter interfacial composition and reaction rates. Together, we reveal a new perspective on multiphase chemistry in the atmosphere. 
\end{abstract}

	\maketitle
\section{INTRODUCTION}
\label{sec:Introduction}
Atmospheric aerosol particles are complex chemical systems often composed of water, ions, and organic molecules in a dizzying variety of combinations and concentrations ~\cite{blanchard1989ejection,cunliffe2013sea,prather2013bringing,burrows2014physically,quinn2015chemistry,cochran2017molecular,bertram2018sea}. These tiny chemical reactors transform both atmospheric gases and dissolved species~\cite{bertram2018sea,Jacob2000heterogeneous,abbatt2012quantifying,brown2012nighttime}. We describe in this Review our efforts to build a molecular perspective of aerosol-mediated reactions that occur across a molecularly detailed interfacial region between the liquid and vapor phases. Flow reactor and gas-liquid scattering experiments reveal fascinating phenomena involving competitive solute-solvent and solute-solute reactions at and near the surface, but they stop short of deducing detailed mechanisms. New theoretical approaches provide a means to map these trajectories with atomistic resolution by wedding the powerful tools of statistical mechanics, machine learning, and accurate many-body potentials to create a general framework for gas-liquid reactions.

In order to ground our discussion, we focus on the chemistry of $\n$ in aqueous aerosols.  We chose $\n$ because its hydrolysis to HNO$_3$ and chlorination to $\cl$ are key aerosol-mediated reactions that influence the oxidizing capacity of the troposphere
~\cite{abbatt2012quantifying,brown2012nighttime,dentener1993reaction,tolbert1988antarctic,mozurkewich1988reaction,stewart2004reactive,bertram2009toward,macintyre2010sensitivity,mcduffie2018heterogeneous,holmes2019role,ha2021effects}. $\n$ is a nighttime reservoir for NO$_2$ and NO$_3$, storing them until daytime when their photolysis leads to production of O$_3$ and OH. Its hydrolysis to HNO$_3$ by aqueous aerosols steals away NO$_2$ and NO$_3$ and thus lowers O$_3$ and OH concentrations, while release of $\cl$, upon photolysis, converts the spectator ion Cl$^-$ into a Cl radical and returns NO$_2$ to the atmosphere. Together, these altered OH and Cl concentrations impact the lifetime of greenhouse gases in the atmosphere ~\cite{macintyre2010sensitivity,holmes2019role, ha2021effects}.
	
The reactivity of $\n$ is particularly appealing to investigate from both mechanistic and interfacial perspectives. $\n$ is sparingly soluble ~\cite{fried1994reaction,schwartz1981solubility,cruzeiro2022uptake}, weakly hydrogen bonding, and undergoes significant charge fluctuations (NO$_2^{\delta+}$NO$_3^{\delta-}  \rightleftharpoons$ NO$_3^{\delta-}$NO$_2^{\delta+}$) that impose both electrophilic and nucleophilic character ~\cite{bianco1999theoretical,mcnamara2000structure,hirshberg2018n2o5,molina2020microscopic}. These properties make the fate of $\n$ beguilingly unpredictable and sensitive to molecular composition, thermodynamic conditions, and motions across aerosol depths from the surface to tens of Angstroms deep. The theoretical and experimental approaches below help to reveal the intricate interplay among solute and solvent molecules that may occur at liquid interfaces. The similarly challenging interfacial chemistry of microdroplets is also beautifully explored by Wilson in this volume ~\cite{wilson2023chemistry}. 
	
We begin this Review by developing equations that relate molecular dynamics to macroscopic observables of aerosol chemistry. We describe a reduced description of the dynamics of a gas molecule like $\n$, through initial adsorption onto an aerosol particle, and its subsequent solvation, diffusion, and eventual loss from evaporation or reaction within the vapor-liquid interfacial region and the bulk liquid. The equations that encode this description are amenable to analysis and parameterization from molecular simulations. With this perspective in mind, we then discuss how we can experimentally confront the complexity of environmental interfaces, such as that of sea spray aerosol. The water-ion-organic-surfactant composition of sea spray generates interfacial nucleophiles and catalysts that lead to competitive reactions like the formation of $\cl$, Br$_2$, and nitration products whose branching ratios can be measured. We conclude with some thoughts for refining our toolkits for understanding and predicting reactivity in complex environments.
 
\section{A MOLECULAR FRAMEWORK FOR GAS UPTAKE}
\label{sec:framework}
\begin{figure}
    \centering
    \includegraphics[
    width=\textwidth
    ]{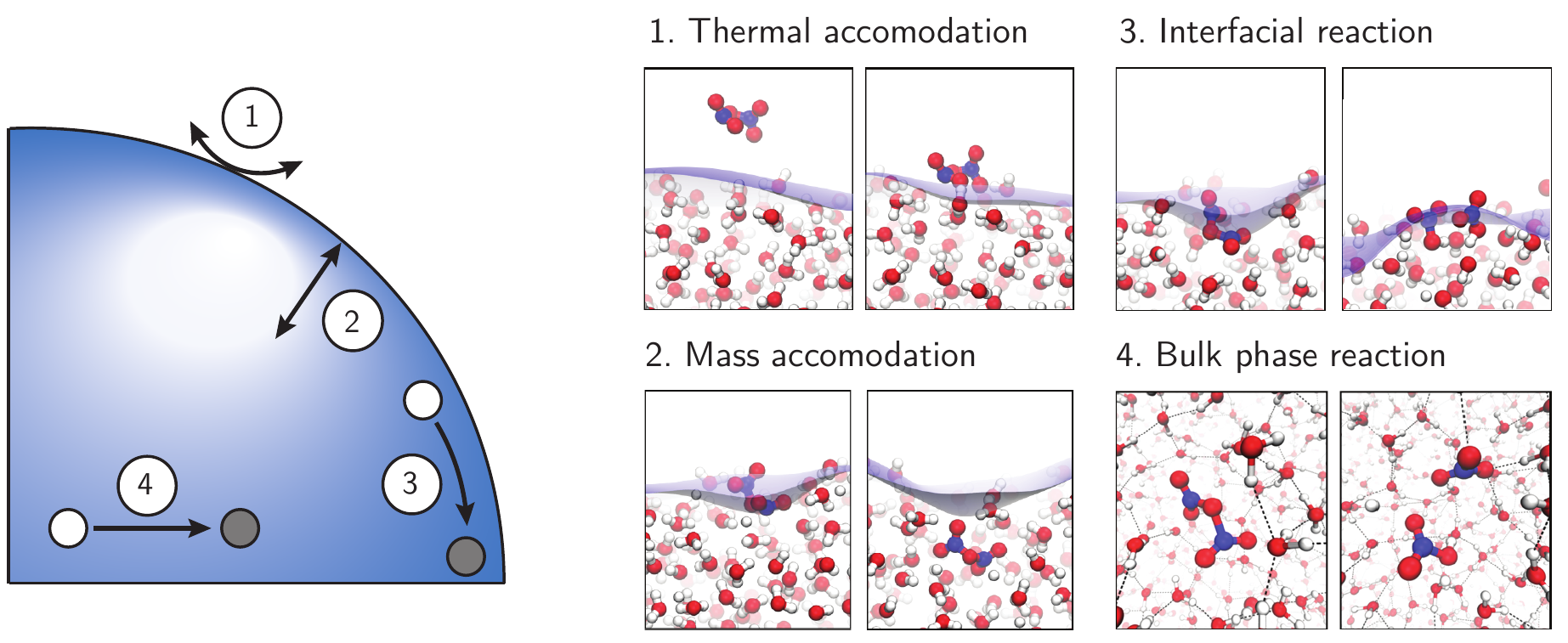}
    \caption{The physical and chemical processes that lead to irreversible loss of a gas particle to an aqueous aerosol, with molecular dynamics snapshots illustrating each for a representative $\n$ gas molecule. Adapted with permission from Reference ~\cite{galib2021reactive}.  Copyright 2021 American Association for the Advancement of Science.}
    \label{fig1}
\end{figure}

Understanding the irreversible loss of a small molecule from the gas phase into aqueous aerosol particles requires that we disentangle molecularly detailed physical and chemical processes in the presence of an extended interface. Early efforts to derive a framework for predicting and explaining trends in gas uptake were largely phenomenological ~\cite{danckwerts1951absorption}. Equations for mass transfer kinetics parameterized by bulk transport properties and constrained by thermodynamic quantities provided an initial formalism and impetus to carefully measure quantities like the Henry's law constant in the laboratory. While the continuum reaction-diffusion equations for processes like those illustrated in Figure \ref{fig1} for $\n$ could be written down and sometimes parameterized, they typically needed to be solved numerically limiting their utility~\cite{sherwood1975mass}. However, simplifying quasi-steady-state assumptions and the linear independence of the individual dynamic processes described by the reaction-diffusion equations admitted an approximation known as the resistor model that is prevalent today in the description of gas uptake~\cite{hanson1994heterogeneous,davidovits2011update,hanson1997surface,worsnop2002chemical}. The resistor model provides explicit analytical expressions for the overall gas uptake probability, is relatively easy to generalize, and is quantitatively accurate in comparison to the full solutions of the reaction diffusion equations. These models, along with advanced formalisms that couple multiple fluxes ~\cite{wilson2023chemistry, shiraiwa2012kinetic, wilson2022kinetic}, build in distinctions between bulk and interfacial processes that are remarkably successful ~\cite{lakey2021kinetic, willis2022coupled} and deployable in chemical transport models ~\cite{riemer2009relative}. However, like their full continuum equivalents, resistor models are phenomenological. They are motivated from underlying approximations to the microscopic dynamics, and consequently lack a means of easily incorporating molecular details that can be important in specific regimes of gas transport like those dominated by near interfacial behavior. 

The presence of an interface between the bulk fluid and the surrounding vapor breaks the symmetry of those two phases, endowing the region in the vicinity of the interface with properties that vary continuously through space ~\cite{hansen2013theory}. For example, the water density profile is not molecularly sharp but varies smoothly between its liquid and vapor values over a nanometer, spread over many molecular diameters by the emergence of capillary waves ~\cite{weeks1977structure,bedeaux1985correlation,willard2010instantaneous}. Other structural properties like water's orientation and polarization, as well as dynamic properties like rotational and translational diffusivity, vary analogously ~\cite{liu2004calculation,liu2005hydrogen,noahvanhoucke2009fluctuations}. This inhomogeneous solvent behavior results in solute properties that also change through space in the vicinity of the air-water interface ~\cite{geissler2013water}. It has been convincingly argued theoretically that there is an affinity of some small molecules and even ions to the liquid-vapor interface, resulting in concentration profiles that are at times enhanced at the interface and relax to their bulk values over molecular lengthscales ~\cite{jungwirth2001molecular, vacha2004adsorption}. Both experimentally and theoretically it has been shown that charged molecules comparable in size to a water molecule are driven to the interface by favorable energies of adsorption, but face unfavorable entropies of adsorption of comparable scale ~\cite{otten2012elucidating, mccaffrey2017mechanism}. The continuous transition from the bulk gas to the bulk fluid through an interface with finite thickness and distinct chemical and physical properties from the two confining phases is not naturally accommodated within a formalism based in the continuum.

In general, we must confront in molecular detail the ways in which properties that shape gas uptake like reaction rates, diffusivities, and concentrations vary in space and how they are coupled dynamically. A perspective for doing so is to integrate out the motion of the solvent and derive an effective equation of motion for the dilute reactive species. If the solvent relaxes quickly, this equation will be a Fokker-Planck equation for the remaining particles ~\cite{zwanzig2001nonequilibrium}. Assuming that we can ignore interparticle correlations and that internal motion is fast relative to translational motion through the fluid, the likelihood of finding a particle of type $i$ at position $\vc{r}_i$ and with velocity $\vc{v}_i$ is $f_i (\vc{r}_i, \vc{v}_i,t)$ and evolves dynamically in time $t$ as	
\begin{equation}
\label{Eq1}
\frac{\partial f_i}{\partial t} = - \vc{v}_i \frac{\partial f_i}{\partial \vc{r}_i} + \frac{1}{m_i} \frac{\partial F_i}{\partial \vc{r}_i}\frac{\partial f_i}{\partial \vc{v}_i}+\frac{\partial }{\partial \vc{v}_i}\frac{\Gamma_i \vc{v}_i f_i}{m_i} + \frac{k_\mathrm{B} T}{m_i^2} \frac{\partial }{\partial \vc{v}_i} \Gamma_i \frac{\partial f_i}{\partial \vc{v}_i} - \sum_j k_{ij}f_i f_j
\end{equation}
where $m_i$ is the molecule's mass, $\Gamma_i (\vc{r}_i )$ is the friction the particle feels due to the other molecules in the liquid and vapor ~\cite{lau2007state}, $F_i (\vc{r}_i )$ is the potential of mean force for moving a tagged particle in space ~\cite{chandler1987introduction}, and $k_{ij} (\vc{r}_i )$ are the bimolecular rates that result in the loss of particle type ~\cite{zwanzig1990rate}. The presence of the explicit velocity dependence in Equation \ref{Eq1} reflects the fact that in the gas phase, molecular center of mass velocities stay correlated over long times. This is not so in the liquid, or even at the interface, where the high densities of surrounding molecules result in high collision rates and fast momentum decorrelation times. The temperature $T$ and Boltzmann's constant $k_\mathrm{B}$ reflect the fact that the quantities that enter this equation are averaged over the other degrees of freedom in the system and are thus state point dependent. This is a single particle Fokker-Planck equation with additional losses through irreversible reactivity affected by bimolecular processes.

\subsection{Parameterizing Molecularly Resolved Mass Transport Equations}
\label{sec:dim_red}

To use Equation \ref{Eq1}, it needs to be parameterized. In particular, the potential of mean force $F_i (\vc{r}_i )$ , friction $\Gamma_i (\vc{r}_i )$, and rate constants  $k_{ij} (\vc{r}_i )$ need to be evaluated. There is little chance that experiments alone can do this directly, though observations in the bulk liquid and vapor phases can provide limiting values far from the interface. Overwhelmingly, molecular dynamics simulations and its extensions have been used for this task. Since the earliest reports of specific ion adsorption to air-water interfaces ~\cite{jungwirth2001molecular}, there have been countless molecular dynamics studies reporting potentials of mean force for small molecule gases ~\cite{tobias2013simulation}. Most studies employ umbrella sampling, in which the potential of mean force is calculated from its relation to a probability ~\cite{frenkel2002understanding}.  Taking an air-water interface translationally invariant parallel to the interface and denoting  as the direction perpendicular to it, the potential of mean force is given by $F_i (z)=-k_B T \ln \langle \delta(z-z_i(t)) \rangle +F_b$, where $F_b$ is a reference value typically taken in the bulk of vapor and $\langle \delta(z-z_i(t)) \rangle$ is the thermally averaged Dirac delta function for particle $i$'s position. 

\begin{figure}
    \centering
    \includegraphics[
    width=\textwidth
    ]{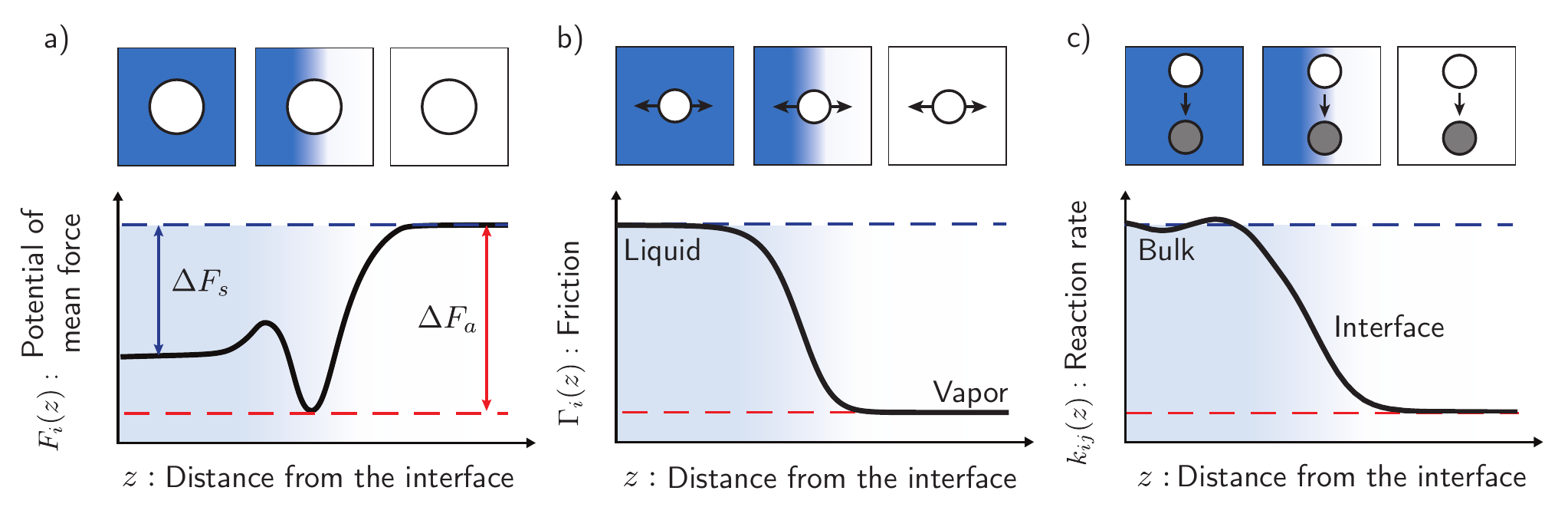}
    \caption{Spatially dependent properties affecting gas uptake from gas phase (depicted in white) to bulk liquid (depicted in blue) via the interfacial region (shaded). a) The potential of mean force for a soluble gas molecule where the vapor value is taken as a reference. The values $\Delta F_s$ and $\Delta F_a$ are the solvation and interfacial adsorption free energies, respectfully. b) The friction for moving a molecule perpendicular to the interface is larger in the liquid than in the vapor. c) The rate for a reaction which is suppressed at the interface, like one which proceeds through significant charge separation, where the molecular lengthscale $\xi$ denotes interfacial width. }
    \label{fig2}
\end{figure}

A characteristic function form for  is shown in Figure \ref{fig2}a. The free energy difference between its limiting values in the liquid and vapor $\Delta F_s$ is the solvation free energy and determines the dimensionless Henry's law constant, $K_H=\exp[-\Delta F_s/k_B T]$. The free energy difference between the vapor and minimum at the interface defines a free energy of adsorption, $\Delta F_a$, which determines the relative equilibrium enhancement or suppression of concentration at the interface relative to the bulk, and the maximum rate of evaporation. The accuracy of $F_i (z)$ and the free energetics derived from it depend sensitively on the molecular model used ~\cite{garrett2006molecular}. The development of many-body force fields that incorporate high level quantum chemistry calculations into their parameterization and concurrent advancements in \emph{ab initio} molecular dynamics simulations have enabled the computation of potentials of mean force accurately enough to make quantitative predictions concerning solubilities and surface affinities ~\cite{cisneros2016modeling, markland2018nuclear}.

It is particularly challenging to develop molecular models that correctly describe both gas and condensed phases, as well as the inhomogeneous interfacial region. A water model that captures the correct physics is MB-pol ~\cite{cisneros2016modeling, babin2013development, babin2014development}, which has been shown to yield quantitative accuracy for molecular properties across the phase diagram of water ~\cite{reddy2016accuracy, bore2023realistic}, including the water-vapor interface ~\cite{muniz2021vapor}. MB-pol is based on the many-body expansion of the total energy and combines an explicit, data-driven representation of short-range, low-order many-body interactions with classical electrostatics, dispersion, and polarization describing long-range and higher-order many-body interactions. This functional form enables the parameterization of many-body potential energy functions from highly accurate electronic structure calculations because the time-consuming reference calculations need to be performed only for the low-order many-body terms. The MB-pol model contains explicit representations for the important two- and three-body terms, while the improved MB-pol(2023) adds a four-body term ~\cite{zhu2023mbpol}. The extension of this modeling framework to mono-atomic ions ~\cite{bajaj2016toward,riera2017toward,paesani2019chemical} and small molecules in water and mixtures ~\cite{riera2020data, bullvulpe2021mbfit} is termed MB-nrg and like MB-pol enables condensed phase simulations at effective coupled cluster accuracy. We recently developed MB-nrg potentials for $\n$ in water ~\cite{cruzeiro2021highly}, which we have employed here to study $\n$ uptake into water ~\cite{cruzeiro2022uptake}.


In comparison to the potential of mean force, the parameterization of spatially varying dynamical quantities is generally less developed. It has been inferred experimentally that relaxational dynamics and reaction rates differ at the air-water interface, but these studies are limited. Surface selective spectroscopy offers a means of making quantitative measurements on ultrafast interfacial dynamics. Observations of a three times shorter hydrogen bonding lifetime at the interface compared to its value in the bulk have been reported ~\cite{hsieh2011ultrafast}. Changes in reaction rates at the air-water interfaces have recently been reported by mass spectrometry of microdroplets, though these measurements are currently difficult to interpret ~\cite{wilson2023chemistry}. Some interfacial halide reactions have been better quantified, for example the Cl$_2^-$. radical attack on ethanol at the air-liquid interface has been measured as proceeding twice faster at the interface relative to the bulk liquid phase ~\cite{strekowski2003direct}. Undoubtedly for ion-dipole reactions that proceed at the diffusion limit in the vapor and much slower in the liquid phase, the interface must offer some distinct value. 

Molecular simulations offer the most reliable means of deducing the spatial distribution of dynamical properties, however there has thus far been much less development along these lines of inquiry. Friction profiles, $\Gamma_i (\vc{r}_i )$, are generically expected to interpolate between large values in the liquid and much smaller values in the vapor like that schematically shown in Figure \ref{fig2}b, with the vapor typically 10$^3-10^4$ times smaller. The profile is expected to roughly follow the solution density ~\cite{liu2004calculation}. Linear response theory admits tractable expressions for computing the spatial dependent friction. For friction opposing motion along the $z$-direction, $\Gamma_i(z) = \int_0^\infty dt \langle \partial_z U(0) \partial_z U(t)\rangle_z /k_\mathrm{B} T$, which is an autocorrelation function of the $z$-component of the force on molecule  from potential $U$ conditioned the molecule being at a position  $z$ ~\cite{chandler1987introduction}. Similar expressions have been developed by Hummer for friction functions along collective variables and offer reasonable alternatives ~\cite{hummer2005position}. The study of the spatial dependence of reaction rate constants computationally is complicated by the need to represent bond making and breaking, and the need to span the timescales available to typical simulations with those of thermally activated reactions. While in principle \emph{ab initio} molecular dynamics simulations can be used for the former ~\cite{car1985unified}, the poor system size scaling makes it difficult to model extended interfaces ~\cite{kuo2004ab}. 

	Modern neural-network based forcefields parameterized using machine learning techniques have opened an exciting frontier for solving this problem ~\cite{niblett2021learning,behler2017first,yang2022using}. To solve the timescale problem, we have developed a trajectory-based importance sampling technique ~\cite{bolhuis2002transition}. Transition Path Sampling + Umbrella Sampling (TPS+U), affords a relatively simple means of computing the rate constant continuously as a function of position relative to a bulk reference ~\cite{schile2019rate} using the side-side correlation function representation,  $\ln k_{ij}(z)/k_{ij} = \ln [\langle h_A(0) h_B(t) \rangle_z/ \langle h_A(0) h_B(t) \rangle] -[F(z)-F_b]/k_\mathrm{B} T$, where $h_x$ is an indicator for being in the $x=A$ reactant or $x=B$ product state. TPS+U has been used to evaluate the rate profiles for mono-molecular and bimolecular processes near liquid-vapor, and liquid-solid interfaces ~\cite{niplett2021ion,singh2022peptide,kattirtzi2017microscopic}. A characteristic example for a reaction requiring significant charge redistribution is shown in  Figure \ref{fig2}c, where due to insufficient screening at the interface the rate is suppressed ~\cite{venkateshwaran2014water,baer2014investigation, wang2009depth}. Condensation reactions or substitution reactions suppressed by polar solvents would likely see this profile reversed, with a rate enhancement observed at the interface ~\cite{reichardt2010solvents, griffith2012situ}.

When parameterized and with appropriate boundary conditions, Equation \ref{Eq1} can describe the processes that dictate gaseous uptake illustrated in Figure \ref{fig1}. For example, in the absence of reactive losses it can describe how an incoming gas molecule initially impinges upon an air-water interface, and the likelihood that it thermalizes its velocity at the interface before evaporating, in a process known as thermal accommodation. Detailed molecular dynamics studies have found universally that for all soluble molecules, the thermal accommodation coefficient is nearly 1 ~\cite{roeselova2003impact,vieceli2004accommodation,morita2004mass,varilly2013water}. This high probability to attach to the interface long enough for the molecule's momentum to decorrelate is the result of the fact that even modestly soluble molecules will have a favorable free energy of adsorption to the interface and thus face a thermodynamically uphill process to evaporate. Provided the molecule is thermally accommodated, its subsequent motion is well described by the over-damped limit of Equation \ref{Eq1} ~\cite{zwanzig2001nonequilibrium},
\begin{equation}
\label{Eq2}
\frac{\partial \rho_i}{\partial t} =\frac{\partial }{\partial \vc{r}_i} \frac{1}{\Gamma_i} \frac{\partial F_i}{\partial \vc{r}_i}\rho_i +\frac{\partial }{\partial \vc{r}_i} \frac{k_\mathrm{B} T}{\Gamma_i} \frac{\partial \rho_i}{\partial \vc{r}_i} - \sum_j k_{ij}f_i f_j
\end{equation}
where
\begin{equation}
\rho_i(\vc{r}_i,t) = \int d\vc{v}_i \, f_i(\vc{r}_i,\vc{v}_i,t)
\end{equation}
is the number density of species $i$ at $\vc{r}_i$. This high friction limit can subsequently model processes like mass accommodation ~\cite{von2020multiscale}, which is the branching ratio between solvation into the bulk of the fluid and evaporation. By simultaneously evolving expressions for multiple species, solute-solute reactions with individual spatially dependent concentration profiles can be modeled. Encoded in $F_i (\vc{r}_i)$ is the effect of barrier molecules like surfactants, and how they might mediate mass transfer into the bulk solution. With a constrained equilibrium initial condition at the interface and sink boundary conditions in the bulk liquid and vapor, the mass accommodation is computable from total accumulated flux into the liquid.  This process is generally not determined solely by the relative activation barriers to move from the interface into the bulk liquid or vapor, but rather by a combination of those thermodynamic factors and the very different values of the friction opposing that motion due to the significant density variations around the interface. Additionally, in the presence of irreversible reactive losses, Equation \ref{Eq2} provides a tractable means of computing the reactive uptake coefficient, $\gamma_i$, the fraction of collisions leading to reaction, with reasonable molecular detail ~\cite{cruzeiro2022uptake}.

\subsection{Reaction Diffusion Length as a Guide for Limiting Behavior}
\label{subsec:CV_sel}
The extent to which molecular details matter for gas uptake depends on the lengthscale over which relevant properties vary, and the depth that a gas molecule likely penetrates into the fluid. Molecular simulation and experimental observations have routinely found that equilibrium concentration profiles vary over a nanometer away from a liquid-vapor interface, reflecting a correlation length that is molecular in character convoluted with the propensity of a free interface to exhibit capillary waves ~\cite{bedeaux1985correlation,willard2010instantaneous,liu2004calculation,liu2005hydrogen, noahvanhoucke2009fluctuations,rowlinson2003molecular,cox2022dielectric,shin2018three}. Inferences on the variation of reaction rates and transport properties are less common, though those few studies similarly find typical variations over a few molecular diameters from the interface. These likely reflect the known variations of dielectric and density fluctuations that vary over molecular lengthscales away from extended interfaces. In the following we will denote this molecular lengthscale  and anticipate it is nanometers in size.

An estimate of the distance a molecule is likely to penetrate the fluid is given by its reaction-diffusion length $\ell_i$. For a small molecule $i$ with diffusion constant $D_i=k_B T/\Gamma_i$, that can react bimolecularly with species $j$ at concentration $\rho_j$, the reaction-diffusion length is
\begin{equation}
\ell_i = \sqrt{\frac{k_\mathrm{B} T/\Gamma_i}{\sum_j k_{ij} \rho_j} }
\end{equation}
where the summation is taken over all potential reactions to account for all irreversible loss channels. While in principle the reaction rates, concentrations, and diffusion constants depend on depth, an approximate length can be arrived at by evaluating each at their bulk values. Values of the reaction-diffusion length vary widely depending on the reaction, external conditions, and solute concentrations. For reactions of relevance to atmospheric chemistry involving dissolved Cl$^-$ ions, reaction diffusion lengths span nanometers for production of $\cl$ from $\n$ or BrCl from HOBr to micrometers for production of Cl$_2$ from HOCl ~\cite{liu2001equilibrium} (at aerosol [Cl$^-$]  5 M and pH  2 ~\cite{bertram2018sea, angle2021acidity}).

	Some simplifications exist that admit analytical predictions to be made and experimental observations interpreted. Typically, reasonable assumptions include a separation of scales between and the size of the aerosol particle, $R>\ell_i$  and, for laboratory experiments, that mass transport limitations from the vapor are not important. As discussed, most molecules will be thermally accommodated on the interface upon impact. Under these assumptions and in the limit that the $\ell_i/\xi \ll 1$, the gas molecule is not likely to penetrate deep into the aerosol. Consequently, spatial gradients are not important, and Equation \ref{Eq2} reduces to
\begin{equation}
\frac{\partial \rho_i}{\partial t} = - \sum_j k_{ij}(z_0) \rho_j(z_0) \rho_i(z_0) \xi 
\end{equation}
which is just the usual form for chemical kinetics in two-dimensions, in which case the relevant rates and concentrations are those averaged over $\xi$ of the liquid-vapor interface, denoted $z_0$. In this case, the reactive uptake coefficient is given by the ratio of the irreversible reactions to the rate of evaporation, $k_e$,
\begin{equation}
\gamma_i = \frac{\sum_j k_{ij}(z_0) \rho_j(z_0)  \xi}{k_e + \sum_j k_{ij}(z_0) \rho_j(z_0)  \xi}
\end{equation}
where $k_e$ can be computed from the free energy of adsorption, $k_e=\sqrt{2k_B T/\pi m_i }  \exp[ \Delta F_a/k_B T]/2$. In the opposite limit that $\ell_i/\xi \gg 1$ assuming the molecule is mass accommodated, the molecule is likely to move far away from the interface before reacting. In such a case, we can neglect the spatial dependence of the potential of mean force, friction, and rates in the interfacial region, and replace them with their bulk values as the bulk phase behavior with its translational invariance will dominate the losses. The Fokker-Planck equation reduces in this limit to
\begin{equation}
\frac{\partial \rho_i}{\partial t} =\frac{k_\mathrm{B} T}{\Gamma_i} \frac{\partial^2 \rho_i}{\partial z^2}  - \sum_j k_{ij}(z_0) \rho_j(z_0) \rho_i(z_0) \xi 
\end{equation}
which is just the usual reaction diffusion equation with bulk phase diffusion constant $D_i=k_\mathrm{B} T/\Gamma_i$. It is known from the resistor model approximation that an accurate estimate is  
\begin{equation}
\label{eq8}
\gamma_i = \sqrt{\frac{2 \pi m_i}{\Gamma_i} \sum_j k_{ij} \rho_j} e^{-\Delta F_s/k_\mathrm{B} T}
\end{equation}
which is a ratio of the total irreversible reaction rate to the effective rate to translate given by the bulk friction, weighted by the equilibrium likelihood of finding the gas phase molecule in the solution ~\cite{davidovits2011update, shiraiwa2012kinetic, worsnop1989temperature}. In between these two limits, when $\ell_i$ is comparable to $\xi$, the detailed, spatially varying properties of the system in the interfacial region must be considered to accurately model the gas uptake. 

\subsection{Uptake of $\n$ on Pure Water as a Canonical Example}
\begin{figure}
    \centering
    \includegraphics[
    width=11cm
    ]{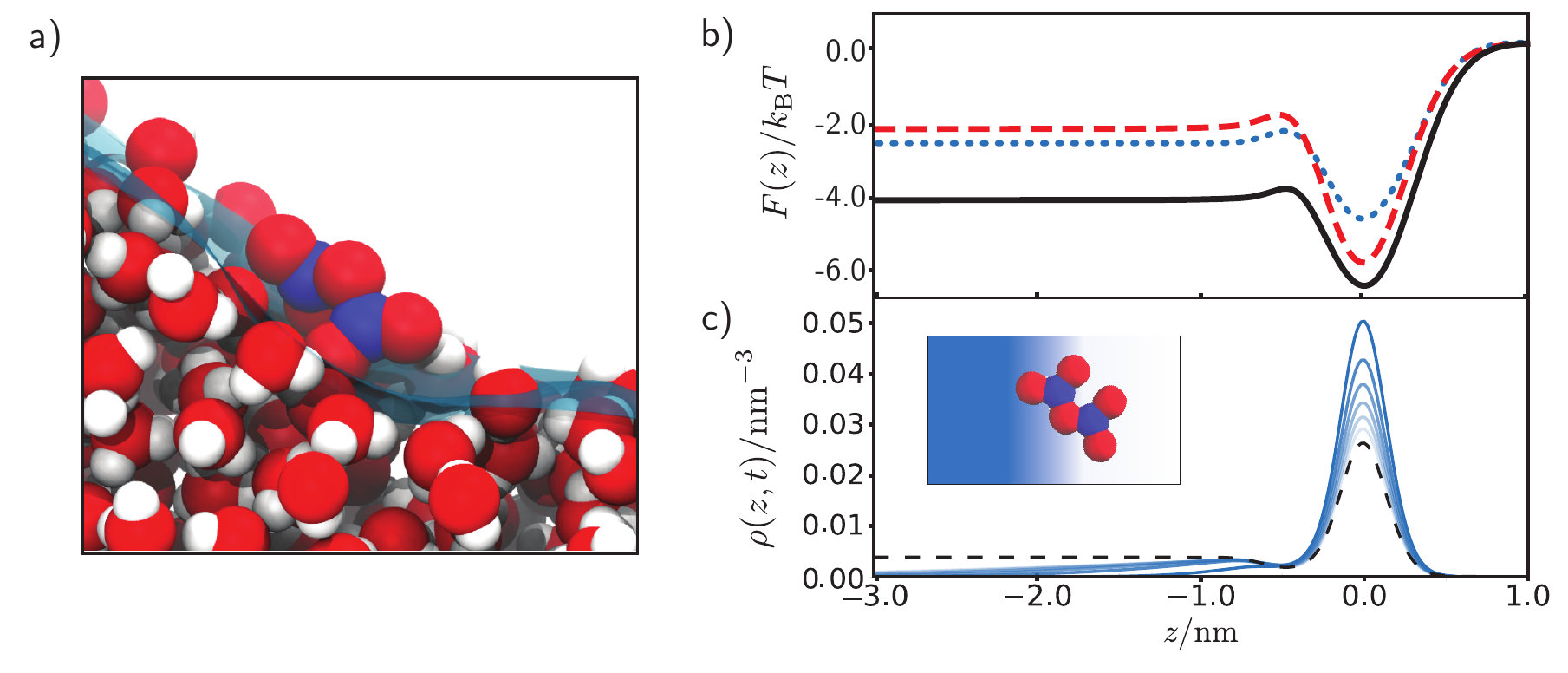}
    \caption{Reactive uptake for $\n$ into a pure water aerosol. (a) A representative snapshot of a molecular dynamics simulation showing an $\n$ molecule and surround water molecules with a representation of the instantaneous liquid-vapor interface. (b) Potentials of mean force computed from an empirical forcefield (blue dotted line), a neural network potential (red dashed line) and a many-body potential (black solid line). (c) Evolution of the $\n$ concentration in the presence of irreversible hydrolysis. The dashed black line is the equilibrium concentration profile in the absence of hydrolysis. Adapted with permission from Reference ~\cite{cruzeiro2022uptake}.  Copyright 2022 Springer Nature.}
    \label{fig3}
\end{figure}

 A canonical case that we have previously applied this framework to is the reactive uptake of  $\n$ through hydrolysis in a pure water droplet illustrated in Figure \ref{fig3}a. The bulk reaction diffusion length was estimated to be between $\ell_i=$15-30 nm, placing it in the regime where molecular details are potentially relevant ~\cite{cruzeiro2022uptake, gaston2016reacto}. However, the fast hydrolysis kinetics restricted careful study of the absolute rate of the reaction, and concurrently, precluded experimental determination of the Henry's law constant. Field measurements taken across a variety of temperature and aerosol concentrations have reported reactive uptake coefficients between 10$^{-4}$ and 0.1 ~\cite{mcduffie2018heterogeneous}, with expectations for pure and salty water aerosol to be between 0.01 and 0.04 ~\cite{stewart2004reactive, bertram2009toward, gaston2016reacto, vandoren1990temperature}.
 
For a pure aqueous aerosol, the only reactive channel for $\n$ is hydrolysis, and under such conditions the rate becomes first order. The resultant Fokker-Planck equation for a flat interface oriented perpendicular to the $z$ axis is
\begin{equation}
\label{Eq9}
\frac{\partial \rho}{\partial t} =\frac{\partial }{\partial z} \frac{1}{\Gamma(z)} \frac{\partial F(z)}{\partial z}\rho(z,t) + \frac{\partial }{\partial z} \frac{k_\mathrm{B} T}{\Gamma(z)} \frac{\rho(z,t)}{\partial z}-  k(z) \rho(z)
\end{equation}
which to solve required that we evaluate the $z$-dependent friction $\Gamma(z)$, potential of mean force $F(z)$, and hydrolysis rate $k(z)$. We developed a series of molecular models with increasing sophistication to compute the nonreactive parameters. Initially a simple point charge model was investigated ~\cite{hirshberg2018n2o5}, with subsequent studies based on a neural network fit to density functional theory, revPBE-D3 ~\cite{galib2021reactive}, and finally an MB-nrg potential ~\cite{cruzeiro2022uptake,cruzeiro2021highly}. The resulting potentials of mean force are shown in Figure \ref{fig3}b, which all found a net driving force for solvation into the liquid, but also deep minima at the interface reflecting the prevalence for specific surface adsorption. While the overall form of  is conserved between the models, differences on the order of 1-2 $k_\mathrm{B} T$ in the solvation and adsorption free energies resulted in predictions for the Henry's law constant that varied by nearly an order of magnitude. The MB-nrg model predicted a Henry's law constant of 3.0$\pm$0.4 M/atm, likely the best value available theoretically or experimentally. These models were much more consistent with the variation of the friction between the bulk and interface, a factor of 3.5 smaller at the interface relative to bulk, and a bulk diffusivity $D=k_\mathrm{B} T/\Gamma =1.9 \times 10^{-5}$  cm$^2$/s. All models predicted a thermal accommodation statistically indistinguishable from unity.

	The parameterization of the $z$-dependent hydrolysis rate is not as amenable to systematic study, theoretical techniques capable of making chemically accurate estimates of liquid phase reaction barriers are not generally available. Nevertheless, using a neural network form, a machine-learning based potential fitted to revPBE-D3 density functional theory was parameterized and used to study $\n$ hydrolysis in the bulk liquid and at the air-water interface ~\cite{galib2021reactive}. Studying an ensemble of reactive trajectories, we found that the mechanism of hydrolysis was largely conserved between the bulk liquid and the air-water interface. In both cases, hydrolysis proceeded through the concerted dissociation of $\n$ and donation of a proton or OH$^-$ group to the transiently formed NO$_3^-$ and NO$_2^+$ moieties. Notably a long-lived NO$_2^+$ was not observed ~\cite{molina2020microscopic}. The rate of hydrolysis in the bulk relative to the interface was 4 times faster, which was rationalized based on the destabilization of the transient charge separation, NO$_2^{\delta+}$NO$_3^{\delta-}$, at the interface ~\cite{hirshberg2018n2o5, galib2021reactive}.

While the trend is likely correct, the absolute rates are subject to errors in the electronic structure. Numerically evaluating Equation  \ref{Eq9} using the established potentials of mean force and friction, we could bound the likely values of the hydrolysis rate by computing the reactive uptake coefficient and comparing to the observed experimental range. The most likely value was found to be $k=4\times 10^7$/s ~\cite{cruzeiro2022uptake}. Shown in Figure \ref{fig3}c is an example of the evolution of the density profile, beginning in a thermally accommodated interfacial state, and evolved in time with a sink boundary condition in the vapor to enforce evaporative loss. The structure in the potential of mean force is evident in the suppression of probability just below the Gibbs dividing surface. Because of the barrier to solvation and corresponding minima at the interface, the propagation length of =$\n$ into the bulk fluid is only around 2 nm, much less than expected from the naive bulk reaction-diffusion length estimation. Given the structure of the potential of mean force, an interpretation of this result is that the diffusion constant is effectively renormalized to a much smaller value. The reactive uptake computed from the numerical solution of Equation \ref{Eq9} is  $\gamma=0.09$, while that from the approximate Equation  \ref{eq8} is $\gamma=0.24$ confirming that the correlation between mass accommodation, reaction and diffusion necessitates a molecular description for understanding the uptake of $\n$.  

\section{ION AND SURFACTANT CONTROL OF INTERFACIAL REACTIONS}
In parallel with these theoretical approaches, we investigated the interfacial behavior of $\n$ experimentally by exposing it to different co-reactants that also populate the surface region. The hydrolysis of $\n$ in aqueous aerosols is often accompanied by reactions with diverse solutes, including the oxidation of Cl$^-$ and Br$^-$ to photolabile $\cl$ and Br$_2$ and the nitration of aromatic and aliphatic alcohols ~\cite{abbatt2012quantifying,brown2012nighttime,chang2011heterogeneous,behnke1997production,mitroo2019cino2,mcnamara2021observation,jahl2021response,schweitzer1998multiphase,finlayson2009reactions,lopez2012temperature,heal2007aqueous,gross2009reactive}.  Sea spray is a natural example of a complex environmental system that is salty, organic-rich, and surfactant-coated ~\cite{bertram2018sea,gill1983organic,donaldson2006influence,mcneill2014surface,ryder2015role}. These components enable us to investigate a multitude of nucleophiles that attack $\n$ with a view toward unraveling this ``ocean labyrinth" of competitive reaction pathways.

\subsection{Salt Water: A Competition between Solute and Solvent}
\begin{figure}
    \centering
    \includegraphics[
    width=\textwidth
    ]{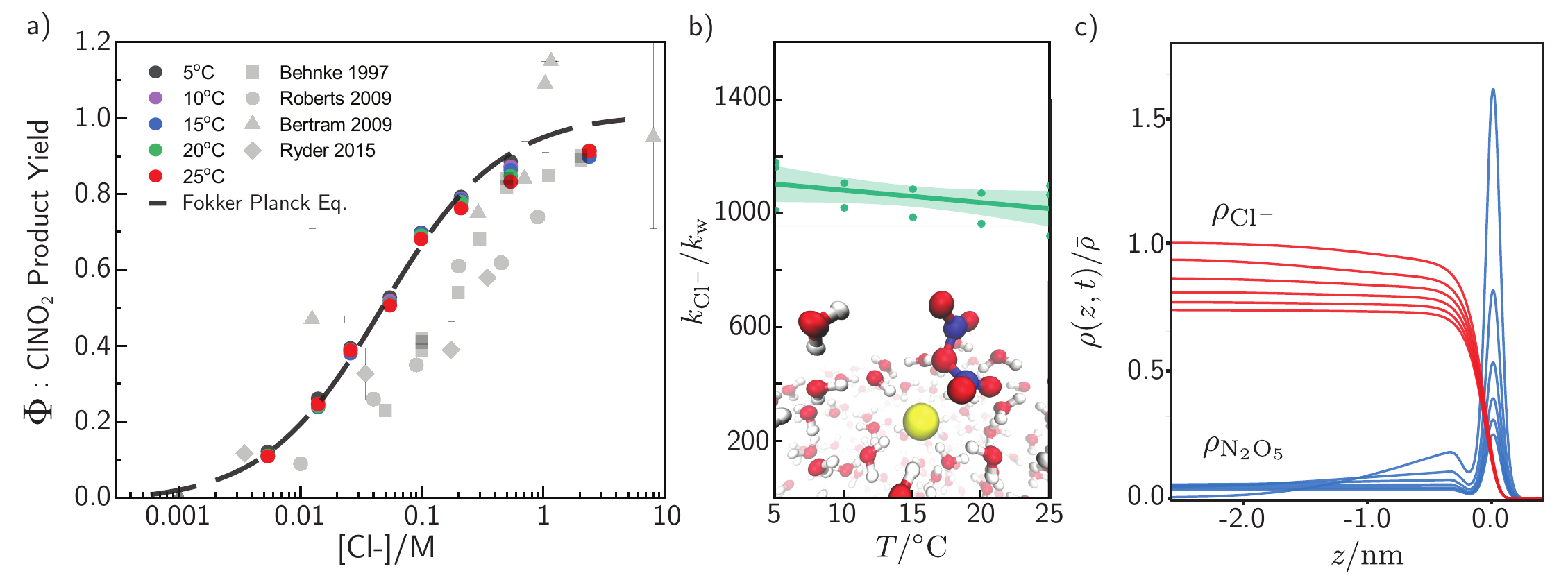}
    \caption{(a) $\cl$ product yield $\Phi$ versus NaCl concentration ~\cite{bertram2009toward,behnke1997production,kregel2023weak,roberts2009laboratory,ryder2015role2} and Fokker-Planck prediction (dashed line) using a bimolecular rate constant ratio  = 1200 for chlorination and hydrolysis. Weakly temperature dependent branching ratio data in D$_2$O are represented by colored circles, literature H$_2$O data are in grey. (b) Inferred  ratio at 0.055 M NaCl. (c) Time dependent solution to Equation \ref{Eq9} for Cl$^-$ and $\n$ densities at $\bar{\rho}$ = 0.05 M. Preliminary studies indicate that the chlorination rate is enhanced and the hydrolysis rate is suppressed in the interfacial region. Panels a and b adapted with permission from Reference ~\cite{kregel2023weak}. Copyright 2019 American Chemical Society.}
       \label{fig4}
\end{figure}

The most abundant reactive ion in seawater and sea spray is Cl$^-$, which varies in concentration from 0.55 M in the ocean to saturation at 6 M in evaporating aerosol particles ~\cite{bertram2018sea}.  These Cl$^-$ ions attack $\n$ to generate $\cl$, a solute-solute reaction that competes with hydrolysis:
\begin{equation}
\label{Eq10}
\n (g) + \mathrm{H}_2 \mathrm{O} (l) \rightleftharpoons 2 \mathrm{NO}_3^- (aq) + 2 \mathrm{H}^+ (aq)
\end{equation}
$$
\Delta H^\circ = -142\, \mathrm{kJ/mol} \quad \Delta S^\circ = -134\, \mathrm{J/mol \,K} \quad \Delta G^\circ = -103 \,\mathrm{kJ/mol} \quad K_\mathrm{eq}(25^\circ \mathrm{C}) = 1\times 10^{18}
$$
\begin{equation}
\label{Eq11}
\n (g) + \mathrm{Cl}^- (aq) \rightleftharpoons \cl (g) + \mathrm{NO}_3^- (aq)
\end{equation}
$$
\Delta H^\circ = -40\, \mathrm{kJ/mol} \quad \Delta S^\circ = 10\, \mathrm{J/mol \,K} \quad \Delta G^\circ = -43 \,\mathrm{kJ/mol} \quad K_\mathrm{eq}(25^\circ \mathrm{C}) = 3\times 10^{7}
$$
Despite the extreme thermodynamic bias toward hydrolysis, chlorination prevails even at concentrations below 0.5 M NaCl ~\cite{kregel2023weak}. This kinetic dominance is shown in Figure \ref{fig4}a, which displays the measured $\cl$ branching fraction versus Cl- concentration for several measurements.  The branching fraction, also called the product yield $\Phi(\cl)$, is given by a quotient of rates  $R_{\mathrm{Cl}^-}/(R_{\mathrm{Cl}^-} + R_w + R_\mathrm{other})$, where $R_{\mathrm{Cl}^-} = k_{\mathrm{Cl}^-} [\mathrm{Cl}^-][\n]$ and $R_w = k_w[\mathrm{H}_2\mathrm{O}][\n]$, and Rather represents other rates, such as nitration discussed below. $\Phi(\cl)$ varies from 0 (no $\cl$ production) to 1 ($\cl$ as the sole product of reacting $\n$) and is roughly 1/2 around 0.2 M NaCl. The extracted ratio of forward rate constants, $k_{\mathrm{Cl}^-}/k_w$, is equal to $(\Phi/(1-\Phi)[\mathrm{H}_2\mathrm{O}]/[\mathrm{Cl}^-]$.  This ratio is close to 1000, strongly favoring chlorination over hydrolysis.  We and others suspected that this propensity for $\cl$ formation must be due to a lower activation energy for Cl$^-$ attack, as it is a better nucleophile than H$_2$O.  To explore this hypothesis, we measured $\Phi(\cl)$ versus temperature from 5 to 25$^\circ$C in solutions ranging from 0.0054 to 0.21 M NaCl ~\cite{kregel2023weak}. The experiments were performed by flowing $\n$ over a flat solution of NaCl in D$_2$O at atmospheric pressure and using iodide chemical ionization mass spectrometry to detect $\cl$ as I($\cl$ )- (D$_2$O was used instead of H$_2$O because of overlap between I(HNO$_3$)(H$_2$O)$^-$ and I($\cl$ )$^-$ signals). The production of $\cl$  over a saturated NaCl solution provides a reference signal for which $\Phi(\cl)$ = 1. These experiments do not measure reactive uptake because of diffusion limitations in the gas phase at atmospheric pressure, but they precisely measure the relative amount of $\cl$  produced at different NaCl concentrations and temperatures. 

	Our data are shown in Figure \ref{fig4}a as narrowly clustered points at each NaCl concentration ~\cite{kregel2023weak}. To our surprise, we find that $\Phi(\cl)$ decreases on average only by 4 $\pm$ 3\% over the 20$^\circ$ C range. An Arrhenius analysis in Figure \ref{fig4}b using $k=A \exp[-E_a/ k_\mathrm{B} T]$ indicates that the activation energy $E_a$  for hydrolysis is only 3.0 $\pm$ 1.5 kJ/mol higher than for chlorination, a difference equal to just 1.2 $k_\mathrm{B}T$ at 25 $^\circ$C.  Instead, the large $k_{\mathrm{Cl}^-}/k_w$ ratio can be attributed to the large ratio of pre-exponential factors,  $A_{\mathrm{Cl}-}/A_w = 300_{(-200)}^{(+400)}$.
	
	Evolution of the overdamped Fokker Planck Equation \ref{fig2} jointly for the  Cl$^-$ and $\n$  concentrations with a bimolecular recombination rate are shown in Figure \ref{fig4}c. As with hydrolysis of $\n$ by pure water, the reactions occur largely within the interfacial region, with the dual reactive channels keeping $\n$ from penetrating more than 1 nm into the liquid (just smaller than 2-5 nm inferred in Ref ~\cite{gaston2016reacto}). The difference in chlorination and hydrolysis appears to arise more from entropic and dynamic factors embedded in A than from the difference in $E_a$. Ongoing theoretical studies suggest that the strongly solvated Cl$^-$ ion must shed its tight water solvation shell upon reaction with the hydrophobic $\n$ to form a nearly as hydrophobic $\cl$, a fluctuation that is easier near the liquid-vapor interface. Implied in this picture is that the transition state
	$[\mathrm{Cl}^{\delta -} \cdots \mathrm{NO}_2^{\delta +} \cdots \mathrm{NO}_3^{\delta -}]^-$
 is entropically favored, and it is enhanced at the interface as the attacking Cl$^-$ loses its ordered solvation shell more than the departing NO$_3^-$ gains one. In contrast, hydrolysis generates HNO$_3$ or H$^+$/NO$_3^-$ from H$_2$O and a hydrophobic $\n$ reactant and so is entropically disfavored as anticipated from the thermodynamic entropy changes in Equations \ref{Eq10} and \ref{Eq11}. This solute-solvent competition is therefore won by the solute Cl$^-$ ion as the surrounding solvent water molecules reposition and reorient to allow Cl$^-$ to approach and react with $\n$ at the N atom of NO$^{\delta+}$.  

The effect of hydration on the mechanisms of hydrolysis and Cl$^-$ substitution of $\n$ has been explored theoretically in small water clusters as model systems, with the earliest work going back over two decades ~\cite{mcnamara2000structure,mcnamara2000exploration}. Recently, reaction products of $\n$ with halides in small water clusters were analyzed via vibrational spectra using a cryogenic photofragmentation mass spectrometry, indicating that $\n$ preferentially undergoes halide substitution over hydrolysis in these model systems ~\cite{kelleher2017trapping}. Later theoretical studies identified key features of the potential energy surface in clusters of $\n$ and Cl$^-$ with one to five water molecules using coupled cluster calculations and density functional theory based \emph{ab initio} molecular dynamics ~\cite{mccaslin2019mechanisms,mccaslin2022effects,gerber2015computational}. It was found that there is almost no barrier for the Cl$^-$ substitution reaction of $\n$ to form $\cl$ when only one water molecule is present. In this case hydrolysis can only proceed through initial formation of $\cl$, which then hydrolyzes.  In contrast, direct hydrolysis requires the presence of at least two water molecules. The structural motifs of the transition states for hydrolysis and Cl$^-$ substitution are conserved upon increasing hydration while the barrier for Cl$^-$ substitution increases monotonically from under 3 kJ/mol to about 12 kJ/mol. Similarly, the reaction barrier for hydrolysis of $\n$ increases from about 25 kJ/mol to about 50 kJ/mol. These results indicate that nucleophilic attack of Cl$^-$ on $\n$ is favored over hydrolysis, as observed experimentally. Once $\cl$ is formed, it can also hydrolyze, however with a larger barrier of about 60 kJ/mol that shows little dependence on the number of water molecules in the cluster. Compared to uncharged clusters containing only $\n$ and water, the presence of Cl$^-$ increases the barrier for hydrolysis substantially, meaning that Cl$^-$ inhibits one-step hydrolysis for clusters of this size, presumably by strengthening the hydrogen bonds between the water molecules. These studies provide a molecular picture of how $\cl$ formation is sensitive to the local aqueous environment, and informs other studies of more complex solutions. 

\subsection{Common Organics and Anions Alter $\cl$ Branching }
\begin{figure}
    \centering
    \includegraphics[
    width=\textwidth
    ]{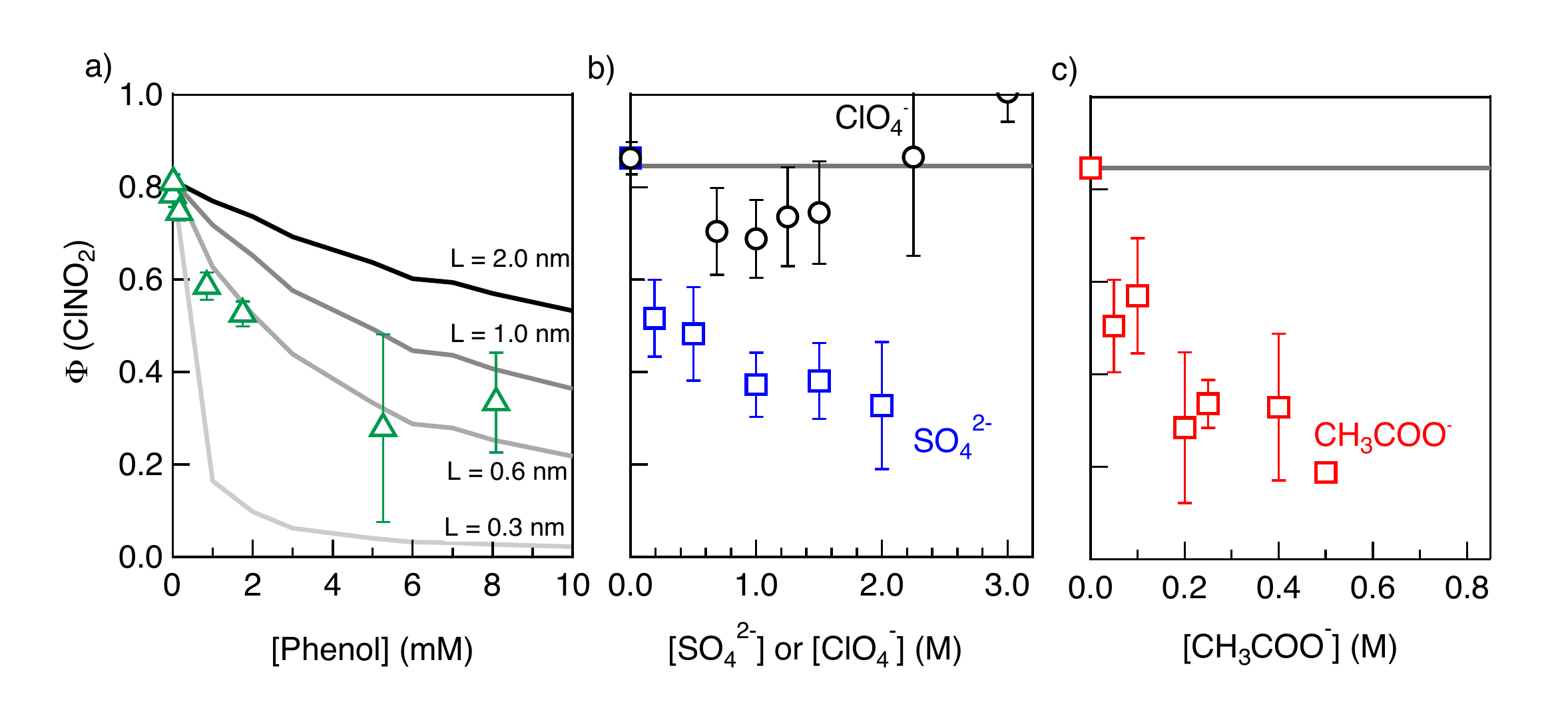}
    \caption{ $\cl$ product yield in a 0.5 M NaCl solution at 209 K upon addition of (a) phenol. Note the small range of concentrations up to just 8 mM. The lines represent model fits for reaction over a depth $L$ described in Ref  ~\cite{mcduffie2018clno2}. (b) sodium sulfate and sodium perchlorate. (c) sodium acetate. Note the different concentration ranges for each salt. Reproduced with permission from References ~\cite{roberts2009laboratory} and ~\cite{mcduffie2018clno2}. Copyright 2015 and 2019 American Chemical Society.}
    \label{fig5}
\end{figure}

Despite the greater reactivity of Cl$^-$ than H$_2$O, our field measurements revealed a startling result: the $\Phi(\cl)$ product yield drops from 0.8 in 0.5 M NaCl to $~\sim$0.2 in seawater samples collected from the Atlantic and Pacific oceans, in controlled blooms in ocean water, and in mimics of the sea surface microlayer ~\cite{ryder2015role}. Recent field studies reveal similar trends ~\cite{wang2017fast, mcduffie2018clno2}.  What does seawater contain that pure salt water does not? Two common classes of species are other ocean anions, such as sulfate and acetate, and aromatic groups found in compounds such as humic acid, chlorophyll, and tyrosine, which we mimic here by phenol. To explore why $\Phi(\cl)$ drops so significantly, we measured it in the presence of these three species using a flow reactor and chemical ionization detection mentioned above. The results for phenol, acetate, and sulfate are shown in Figure \ref{fig5}. Each species indeed impedes $\cl$ production, while the reference ``inert" perchlorate ion changes $\Phi(\cl)$ much less.

We first investigated phenol as a proxy for aromatic-bearing groups in the ocean ~\cite{ryder2015role2}. Phenol may undergo electrophilic attack by $\n$ to produce nitrophenol:  
\begin{equation}
\n (g) + \mathrm{C}_6\mathrm{H}_5\mathrm{OH} (aq) \rightarrow \mathrm{NO}_2\mathrm{C}_6\mathrm{H}_4\mathrm{OH} (aq) + \mathrm{NO}_3^- (aq) 
\end{equation}
This nitration reaction is hundreds of times faster than hydrolysis, and perhaps close to the reaction rate of $\n$ + Cl$^-$ ~\cite{heal2007aqueous}. Phenol is surface-active and requires only millimolar concentrations to populate the interfacial region: for 8 mM phenol in 500 mM NaCl, the Gibbs relative surface excess is +2$\times 10^{13}/$cm$^2$ (excess over the bulk when integrated over the interfacial depth)~\cite{ryder2015role2}. In contrast, NaCl in water raises the surface tension, and the surface depletion of Cl$^-$ is -1$\times 10^{13}/$cm$^2$~\cite{pegram2007hofmeister}.  A continuum reaction-diffusion model incorporating surface adsorption predicts the observed $\Phi(\cl)$ from 0 to 8 mM phenol when hydrolysis, nitration, and chlorination occur within 10 \AA \, of the surface. In this sense, the near-interfacial region becomes a ``battleground" for competing solute-solvent and solute-solute reactions involving $\n$, water, ions, and surface-active species in which $\ell$ and $\xi$ are similar depths.

	In addition to nitration of aromatic rings, $\n$ can also nitrate alcohols ~\cite{gross2009reactive}.  We have indeed observed mono- and di-nitration of pure glycerol, CH$_2$OH(CHOH)CH$_2$OH ~\cite{shaloski2017reactions}. This solvent nitration, the analog of water hydrolysis to HNO$_3$, dominates over the reaction of $\n$ with Br$^-$ to produce Br$_2$ even in 2.7 M NaBr. Thus, $\n$-initiated halogenation does not always win against reaction with a protic solvent. 

	The decrease in $\Phi(\cl)$ by sulfate and acetate, shown in Figure \ref{fig5}, puzzled us more ~\cite{staudt2019sulfate}. The absence of a significant change by ClO$_4^-$ might rule out a dominant role for charge interference (ionic strength effects) in this ion-neutral molecule reaction. One alternative explanation was suggested by \emph{ab initio} simulations by Karimova and Gerber: sulfate and acetate are even better nucleophiles than chloride ~\cite{karimova2020sn2}! They calculate lower energy reaction paths for S$_N$2 attack on NO$_2^{\delta +}$ NO$_3^{\delta -}$ by SO$_4^{2-}$ and CH$_3$COO$^{-}$ than by Cl$^-$:
\begin{equation}
\n (g) + \mathrm{SO}_4^{2-}(aq) \rightarrow  [\mathrm{NO}_2\mathrm{SO}_4]^{-} (aq) + \mathrm{NO}_3^- (aq) 
\end{equation}
\begin{equation}
 [\mathrm{NO}_2\mathrm{SO}_4]^{-} (aq) + \mathrm{H}_2\mathrm{O} \rightarrow \mathrm{SO}_4^{2-}(aq) + \mathrm{NO}_3^- (aq) + \mathrm{H}^+(aq)
\end{equation}
where  $[\mathrm{NO}_2 \mathrm{SO}_4]^-$ is a newly postulated intermediate and  $\mathrm{SO}_4^{2-}$ is regenerated in the second step. These reactions are examples of a powerful motif in seawater and aerosol chemistry: ``spectator" ions such as  $\mathrm{SO}_4^{2-}$ or CH$_3$COO$^-$ can act as catalysts, outcompeting Cl$^-$ and diverting $\n$ toward hydrolysis. From a macroscopic perspective, sodium acetate is mildly surface active (CH$_3$COO$^-$ surface excess of +2$\times 10^{13}/$cm2 at 0.5 M CH$_3$COONa) ~\cite{minofar2007propensity}, while sodium sulfate is surface inactive ($\mathrm{SO}_4^{2-}$ surface excess of -2$\times 10^{13}/$cm$^2$ at 0.5 M Na$_2$SO$_4$)~\cite{pegram2007hofmeister}.  Molecular dynamics and spectroscopic studies further reveal that acetate populates the outermost region ~\cite{minofar2007propensity} and sulfate is depleted over two or more water layers ~\cite{gopalakrishnan2005air}. This interfacial depletion may be one reason that higher concentrations of sulfate than acetate are required to lower $\Phi(\cl)$ in near-interfacial reactions of $\n$.

	Intriguingly, the acetate results imply that longer chain carboxylate surfactants could also be significant inhibitors of $\n$ chlorination, not only because they may physically block $\n$ transport across the surface, but also because they promote hydrolysis over chlorination. As discussed below, these anionic surfactants also repel Cl$^-$ from the interfacial region because of their like charges, providing a third mechanism for reducing $\cl$ production.

\subsection{Soluble Surfactants as Catalysts that Drag and Repel Halide Ions from the Interface}
\begin{figure}
    \centering
    \includegraphics[
    width=12.cm
    ]{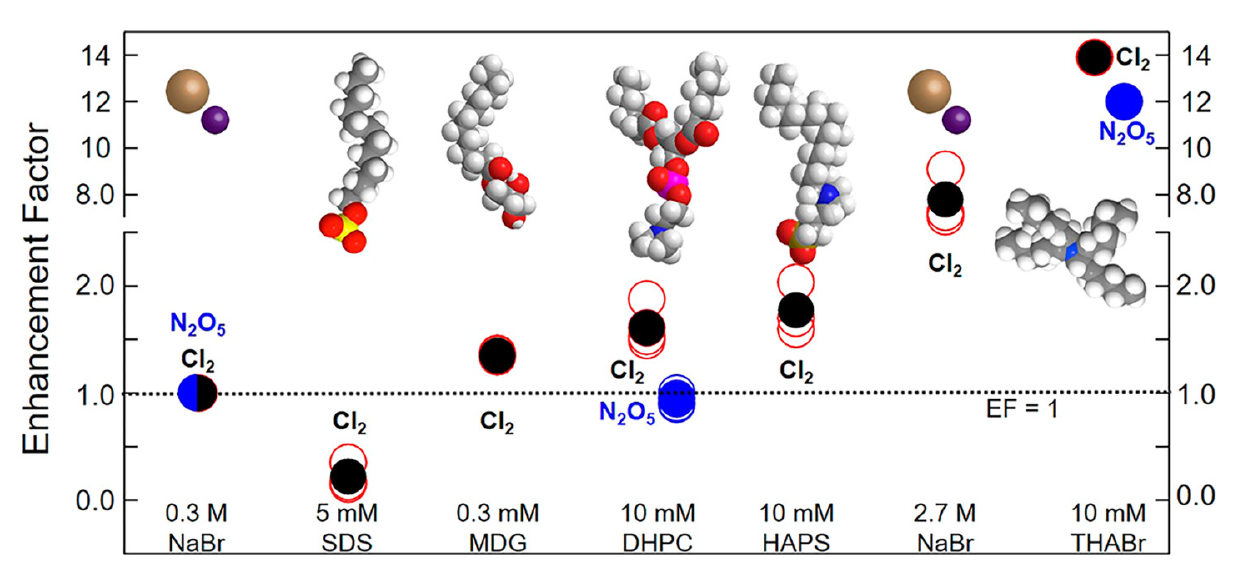}
    \caption{ Reaction rates for the oxidation of Br$^-$ to Br$_2$ by Cl$_2$ (black) or $\n$ (blue) relative to pure 0.3 M NaBr in glycerol.  The surfactants and their fractional surface coverages are sodium dodecyl sulfate (SDS, 53\%), tetrahexylammonium bromide (THABr, 73\%), dihexanoylphoshotidylchloline (DHPC, 61\%), and N-hexadecyl-N,N-dimethyl- 3-ammonio-1-propanesulfonate (HAPS, 69\%). Enhancement factors $>$ 1 indicate that the surfactant catalyzes Br$_2$ production with respect to 0.3 M NaBr/glycerol alone. Reproduced with permission from Reference ~\cite{gord2018control}. Copyright 2019 American Chemical Society.}
    \label{fig6}
\end{figure}

We next used gas-liquid scattering experiments in vacuum to explore the roles of surfactants in directing interfacial reactions ~\cite{faust2016microjets}. These experiments isolate reactions occurring in single gas-surface collisions by avoiding encounters in the air above the liquid. They provide mechanistic insights that complement kinetics from flow reactor studies.  We started our experiments with liquid glycerol because of its low vapor pressure (10$^{-4}$ Torr) in comparison with water (24 Torr) in order to suppress gas-vapor collisions.  We also switched from Cl$^-$ to Br$^-$ because the scattering experiments use a mass spectrometer detector that relies on electron impact ionization, which cannot distinguish between $\n$, $\cl$, and HNO$_3$, as each dissociatively ionizes to NO$_2^+$.  We then simplified the reaction further by replacing  $\n$ with Cl$_2$ because of its higher reaction probability (70\% for Cl$_2$ in glycerol) ~\cite{dempsey2012nearinterfacial}:
\begin{equation}
\mathrm{Cl}_2 (g) + 2 \mathrm{Br}^- (\mathrm{gly}) \rightarrow \mathrm{Br}_2 (g) + 2\mathrm{Cl}^- (\mathrm{gly})
\end{equation}
via a mechanism involving sequential attack of Br$^-$ and loss of Cl$^-$, followed by release of Br$_2$ into the gas phase
\begin{equation}
\mathrm{Cl}_2 (g) \rightarrow \mathrm{Cl}_2\mathrm{Br}^- \rightarrow \mathrm{ClBr}_2^- \rightarrow \mathrm{Br}_3^-
\end{equation}
\begin{equation}
\mathrm{Br}_3^- \leftrightarrow \mathrm{Br}^- + \mathrm{Br}_2 \rightarrow \mathrm{Br}_2 (g) 
\end{equation}
Our objective was to learn how this halogen exchange reaction can be controlled by soluble ionic and zwitterionic surfactants. We chose tetrahexylammonium bromide  (THA$^+$/Br$^-$), sodium dodecyl sulfate (Na$^+$/DS$^-$), and two zwitterions combining positive ammonium and negative phosphate or sulfonate groups ~\cite{faust2013surfactant, gord2018control}.  The coated-wheel scattering experiments do not measure branching fractions but rather the relative production of Br$_2$ (the product of branching fraction and reactive uptake). The results are shown in Figure \ref{fig6}: when compared with no surfactant in a 0.3 M NaBr solution, THA$^+$ increases  Br$_2$ production by 14-fold and DS- decreases it by 5-fold.  Even the zwitterions are mildly catalytic, increasing Br$_2$ production by about 70\%. These results suggest that THA$^+$ drags Br$^-$  to the surface, increasing its interfacial concentration and transforming it into a surfactant anion, while DS$^-$ repels Br$^-$ and depletes the interfacial region. The interplay of positive and negative charges in the zwitterions is more subtle, and simulations indicate that Na$^+$ gathers near the phosphate group while Br$^-$ gathers near the ammonium group ~\cite{vacha2009effects}. 

	It is remarkable that surfactants can enhance gas-liquid reactivity, as they take up space at the surface and should physically block some gas entry into the interfacial region ~\cite{shiraiwa2012kinetic, donaldson2006influence,thornton2005n2o5,mcneill2006effect,	badger2006reactive}. Gases may in fact be more soluble in an organic film than in water ~\cite{clifford2007direct}, while incompletely packed monolayers of soluble surfactants can be porous enough that an interfacial increase in oppositely-charged co-reactant may overwhelm this barrier effect.  The surfactant molecule itself may also be reactive at low surface concentrations but impose barriers at higher coverages ~\cite{burden2014entry}.	 

	The enhanced reactivity of Cl$_2$ in THA$^+$/Br$^-$ glycerol solutions is accompanied by a dramatic increase in Br$_2$ solvation time, from less than 1 $\mu$s in 0.03 M NaBr to greater than 100,000 $\mu$s in 0.03 M THABr ~\cite{faust2013surfactant}. This lifetime is longer because Br$_2$ reacts reversibly with Br$^-$ to form the Br$_3^-$ intermediate (Equation 17) ~\cite{wang1994equilibrium}. The tribromide anion complexes with THA$^+$, perhaps even at the surface. Indeed, a similar complex, tetrabutylammonium tribromide, is a commercial reagent for safely introducing Br$_2$ in bromination reactions.

	This surfactant effect is not limited to Cl$_2$: when $\n$ is used instead, the addition of THA$^+$ increases Br$_2$ production by 12-fold (similar to 14-fold for Cl$_2$) as shown in Figure \ref{fig6} ~\cite{shaloski2017reactions, gord2018control}:
\begin{equation}
\n (g) + 2 \mathrm{Br}^- (\mathrm{gly}) \rightarrow \mathrm{Br}_2 (g) + \mathrm{NO}_3^- (\mathrm{gly}) + \mathrm{NO}_2^- (\mathrm{gly})
\end{equation}
likely via a BrNO$_2$ intermediate that is rapidly attacked by Br$^-$.  Collectively, these studies reveal that cationic surfactants catalytically enhance bromination by enriching the interfacial regions with Br$^-$, while anionic ones repel Br$^-$. In this way, ionic surfactants naturally tilt reactions toward or away from the interfacial region.

\begin{figure}
    \centering
    \includegraphics[
    width=\textwidth
    ]{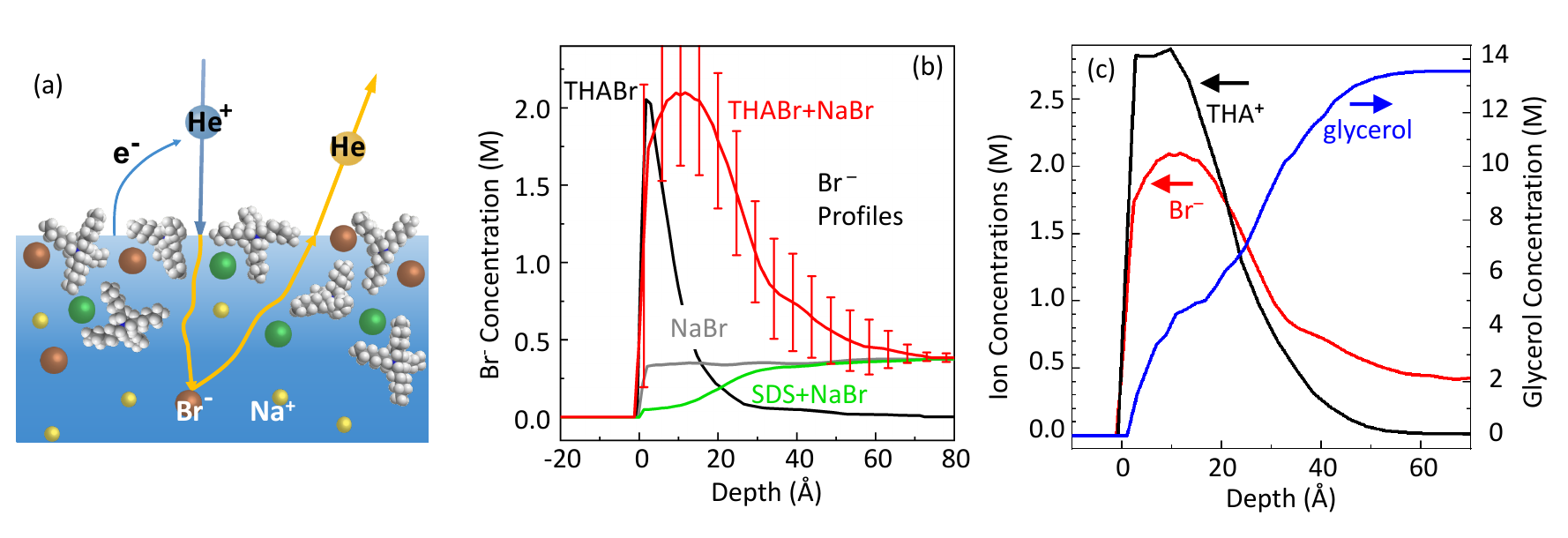}
    \caption{ (a) He$^+$ is neutralized and then traverses the liquid, losing energy and scattering backward from a single atom.  Concentration depth profiles for (b)  Br$^-$ in 0.38 M NaBr, 0.013 M THABr + 0.38 M NaBr, and 0.0065 M SDS + 0.38 M NaBr and (c) comparing Br-, THA$^+$, and solvent glycerol.  Reproduced with permission from Ref ~\cite{zhao2020experimental}. Copyright 2020 American Chemical Society.}
    \label{fig7}
\end{figure}

\subsection{Mapping Solute and Solvent Depth Profiles Angstrom by Angstrom through the Surface}
The diverse effects of ionic surfactants on $\n$ + Br$^-$ reactivity beg the question: just how much do these surfactants alter Br$^-$ interfacial concentrations, and over what depth? To address these questions, we performed helium ion scattering experiments in the laboratory of Gunther Andersson at Flinders University.  This depth profiling technique measures the energy lost by an impinging He$^+$ ion as it is first neutralized at the surface and then undergoes small angle scattering upon penetrating into solution (Figure \ref{fig7}a) ~\cite{andersson2014ion}. A single backward collision with a solute or solvent atom along its trajectory redirects the He atom back through solution into the gas phase, where its kinetic energy is measured. The overall energy loss yields the depth at which the backward collision occurs and the identity of the atom that is struck. These measurements enable Angstrom-scale depths to be determined by comparison with He$^+$ collisions with self-assembled monolayers.

	The resulting Br- depth profiles are shown in Figure \ref{fig7}b for glycerol solutions similar to those in Figure \ref{fig6}: pure NaBr, pure THABr, THABr + NaBr, and NaDS + NaBr ~\cite{zhao2020experimental}. The Br$^-$ distribution is flat for NaBr alone, although small oscillations have recently been measured.~\cite{Kumar2022ion} In contrast, Br$^-$ in pure THA$^+$/Br$^-$ segregates sharply over a width of only 10\AA, about the size of one THA$^+$ or two glycerol molecules. This narrow Br$^-$ distribution is 150 times more concentrated than in the bulk.  In the NaBr + THABr mixture, both Br$^-$ and THA$^+$ are ``salted out", generating a three-fold stronger and broader ion distribution over 30 \AA \, and demonstrating that Br$^-$ and THA$^+$ spread out over at least six glycerol monolayers. Salting-out does not pack more ions into the outermost layer, but rather distributes them over a thicker interfacial region. These measurements demonstrate that THA$^+$ indeed pulls Br$^-$ to the near-surface region, making them available for reaction.  Figure \ref{fig7}c further illustrates that, although solvent glycerol is depleted in the Br$^-$ and THA$^+$ rich interfacial region, it is still present and can participate in reaction.

	Figure \ref{fig7}b, in contrast, shows that the negatively charged dodecyl sulfate anion pushes Br$^-$ ions deeper into solution, significantly depleting Br- over a 40 \AA \, range ~\cite{zhao2020experimental}.  Overall, the Br$^-$ concentration over the top 10 \AA \, varies by 30-fold between the THABr and SDS solutions, similar to the 70-fold change in reactivity of Cl$_2$ with Br$^-$ measured in the reactive scattering experiments.  Slightly weaker depth profiles were measured when NaBr was replaced by NaCl, implying that surfactant-induced interfacial ion enhancements may extend to chloride in seawater and sea spray as well ~\cite{zhao2020competing}).  

\subsection{Bringing the Ocean into the Vacuum Chamber: Gas-Microjet Scattering Studies of $\n$ with Salty and Surfactant-Coated Water}
The transition from low vapor pressure glycerol to high vapor pressure water requires new vacuum-based tools for scattering experiments. Water microjets ~\cite{faubel1988molecular} provide the opportunity to explore reactions of gases with aqueous solutions in the absence of gas-vapor collisions ~\cite{gao2022exploring}. The small surface area of these microjets limits the water vapor density in the surrounding vapor cloud, such that incoming and outgoing gas molecules do not collide with evaporating water molecules. Our scattering apparatus uses a 35 $\mu$m diameter jet of 5 M LiBr traveling at 30 m/s, where the vapor pressure is 2 Torr at 263 K. LiBr is used instead of NaBr because it lowers the solution freezing point. The high jet speed limits the observation time of a 3 mm jet segment to 100 $\mu$s.
	
	Our first experiments again utilized $\n$ + 2 Br$^- \rightarrow $ Br$_2$ + NO$_3^-$ + NO$_2^-$, this time in salty water as a proxy for sea spray ~\cite{sobyra2019production}.  Evaporating Br$_2$ was detected from the 5 M LiBr microjet when it was exposed to $\n$.  We then dissolved the surfactant tetrabutylammonium bromide (TBABr) into the salty microjet solution. TBABr is two CH$_2$ groups shorter than THABr and was used because of its higher solubility in salty water. Inert gas scattering experiments show that the surfactant is at 60\% of full surface coverage at the surface of the jet. To our amazement, the addition of TBA$^+$ eliminated Br2 production in the microjet despite its longer cousin enhancing production by 12-fold in glycerol. We should not have been surprised, however: the lifetime of  Br$_3^-$ (which stores  Br$_2$, Equation 17), was extended from $<10 \mu$s to $>100,000 \mu$s in glycerol upon addition of THA$^+$/Br$^-$ ~\cite{faust2013surfactant}. We estimate a similar lifetime increase in LiBr/H2O, making the TBA$^+$/Br$_3^-$ complex last much longer than the 100 $\mu$s observation time of the microjet. In this sense, the disappearance of the Br2 signal supports the idea that cationic surfactants can significantly increase the lifetime of Br$_3^-$, and thus  Br$_2$. Separate measurements of O$_3$ + Br$^-$ indeed demonstrate enhanced O3 uptake and a longer [BrOOO]$^-$ lifetime when TBA$^+$ is present ~\cite{chen2021impact}. Surfactants may therefore control product lifetimes as well as interfacial reaction rates.

\section{SUMMARY AND FUTURE OUTLOOK }
Our perspectives on aerosol-mediated chemistry summarized here are centered on the importance of viewing processes molecularly, and in particular understanding that the granularity of matter necessitates that we include in our theoretical descriptions and analysis of experimental data a description of the interface between two phases as its own unique environment. Due to their marginal solubility and high reactivity, molecules like $\n$ survive on aerosol particles only long enough to sample the surface and near-surface regions. As we have shown here, the interfacial environment is endowed with physical and chemical properties that can be characterized and measured, and rigorously connected to the concentrations and fluctuations of the molecules that make it up. 
	
	Aerosol droplets contain a diverse range of inorganic ions and organic molecules that make them particularly fascinating. Our studies of even simple aerosol proxies containing salts and surfactants point to the key role of interfacial populations in directing competitive solute-solute and solute-solvent reactions. With significant advances in the preparation and interrogation of heterogeneous systems, these properties and reactions can be measured. The continued advancement of experimental techniques to probe the depth-dependent composition and reactivity of liquid-vapor interfaces, and the molecular models with which to help interpret them, is a grand challenge, but one that will deepen our understanding of the basic physical chemistry of our atmosphere. Our perspective has been made possible by the close collaboration of a team of scientists with complementary expertise, brought together by a center to freely exchange ideas on the basic science of multiphase chemistry in complex environments. 
	
	The concepts and experimental designs reviewed here have largely been motivated by the desire to understand naturally occurring, atmospheric processes. However, they are much more general than this. In the future, elucidating the manner in which molecular concentrations vary near an interface or how reactivity is suppressed or enhanced in strongly heterogeneous media will no doubt enrich related fields of electrochemistry, wastewater treatment, gas scrubbing, separations, sonochemistry, plasma-liquid interactions, and the burgeoning areas of microdroplet and indoor chemistry.

\section*{DISCLOSURE STATEMENT}
\label{sec_disclosure}
The authors are not aware of any affiliations, memberships, funding, or financial holdings that
might be perceived as affecting the objectivity of this review.

\section*{ACKNOWLEDGMENTS}
\label{sec_acknowledgment}
We gratefully acknowledge NSF support for bringing us together through the Center for Aerosol Impacts on Chemistry of the Environment CAICE), a Center for Chemical Innovation, under grant CHE-1801971. We also thank our fellow CAICE researchers whose insights and dedication have made these joint theoretical and experimental studies possible: Vinicius Cruzeiro, Thomas Derrah, Mirza Galib, Joseph Gord, Natalia Karimova, Steven Kregel, Seokjin Moon, Olivia Ryder, Michael Shaloski, Thomas Sobyra, Sean Staudt, and Xianyuan Zhao. Gunther Andersson generously collaborated with us, and Benny Gerber and Mark Johnson have richly improved the perspectives expressed here.

\bibliographystyle{ar-style3}
\bibliography{ref}

\end{document}